\documentclass[aps,prl,twocolumn,superscriptaddress,preprintnumbers]{revtex4}

\usepackage{epsf,epsfig,amsmath,amssymb,amsfonts}
\usepackage{color}
\usepackage{slashed}
\usepackage{hyperref}

\newsavebox{\ns}
\newsavebox{\dbrane}

\def\be{\begin{equation}}
\def\ee{\end{equation}}
\def\bea{\begin{eqnarray}}
\def\eea{\end{eqnarray}}

\def\Dslash{\,\,{\raise.15ex\hbox{/}\mkern-12mu D}}
\def\Dbarslash{\,\,{\raise.15ex\hbox{/}\mkern-12mu {\bar D}}}
\def\delslash{\,\,{\raise.15ex\hbox{/}\mkern-9mu \partial}}
\def\delbarslash{\,\,{\raise.15ex\hbox{/}\mkern-9mu {\bar\partial}}}
\def\pslash{\,\,{\raise.15ex\hbox{/}\mkern-9mu p}}
\def\calDslash{\,\,{\raise.15ex\hbox{/}\mkern-12mu {\cal D}}}

\allowdisplaybreaks

\begin{document}

\title{de Sitter Swampland, $H_0$ tension \& observation}


\vskip 1 cm

 \author{Eoin \'O Colg\'ain}
 \affiliation{Asia Pacific Center for Theoretical Physics, Postech, Pohang 37673, Korea}
 \author{Maurice H. P. M. van Putten}
 \affiliation{Physics and Astronomy, Sejong University, Seoul, Korea}
\author{Hossein Yavartanoo}
\affiliation{State Key Laboratory of Theoretical Physics, Institute of Theoretical Physics, Chinese Academy of Sciences, Beijing 100190, China}

\begin{abstract}

Realising de Sitter vacua in string theory is challenging. For this reason it has been conjectured that de Sitter vacua inhabit the Swampland of inconsistent low-energy effective theories coupled to gravity. Since de Sitter is an attractor for $\Lambda$CDM, the conjecture calls $\Lambda$CDM into question. Reality appears sympathetic to this idea as local measurements of the Hubble constant $H_0$ are also at odds with $\Lambda$CDM analysis of Planck data. This tension suggests that the de Sitter state is unstable, thereby implying a turning point in the Hubble parameter. We present a model relieving this tension, which predicts a turning at small positive redshift $z_*$ that is dictated by present-day matter density $\omega_m$. This feature is easily identified by homogeneous surveys covering redshifts $z \leq 0.1$. We comment on the implications for the Swampland program.  
\noindent 

\end{abstract}

\maketitle

\setcounter{equation}{0}


{\em Introduction -} 
String theory enjoys a billing as the leading candidate for a theory of quantum gravity. It is well-documented that the theory permits a rich landscape of vacua, so much so that de Sitter vacua may be plentiful. However, the theory appears to secretly abhor de Sitter and each manifestation is subject to some criticism (see \cite{McOrist:2012yc,Sethi:2017phn,Danielsson:2018ztv}). The lack of acceptable de Sitter vacua has prompted the conjecture \cite{Obied:2018sgi} that de Sitter belongs to inconsistent low-energy effective theories coupled to gravity, which are deemed to inhabit the Swampland \cite{Vafa:2005ui, Brennan:2017rbf}. This conjecture has attracted considerable attention \cite{Agrawal:2018own,Dvali:2018fqu, Andriot:2018wzk,Achucarro:2018vey,Garg:2018reu,Kehagias:2018uem, Dias:2018ngv, Lehners:2018vgi, Denef:2018etk}. 

Concretely, the conjecture consists of two criteria. The first states that theories with large scalar field excursions in reduced Planck units $M_p = (8 \pi G)^{-1/2}$ are inconsistent, while the second affirms that the gradient of the potential $V$ of a canonically normalised scalar field in a consistent gravity theory must satisfy the bound,  
\be
\label{criteria2}
M_p  |\nabla_{\phi} V|  \geq V c, 
\ee
where $c$ is a constant of order unity, $c \sim O(1)$. Quintessence models naturally satisfy this bound \cite{Agrawal:2018own}. 

The constraint (\ref{criteria2}) is intriguing. First and foremost, it precludes de Sitter vacua where $\nabla_{\phi} V =0$, thus ruling out $\Lambda$CDM even when $c \ll 1$. Secondly, (\ref{criteria2}) represents a marriage of the requirements of early inflationary and late-time dark energy cosmology. In the wake of Planck, potentials are required to be flat, i. e. $\nabla_{\phi} V$ small, whereas we also require that any cosmological constant, in other words, the potential, at late times also be small. In principle, this requires double fine-tuning, but this can be avoided if the potential and its gradient are related. 

Recalling that one realisation of de Sitter is an asymptotic future attractor for the $\Lambda$CDM cosmological model, and $\Lambda$CDM captures beautifully measurements of the Cosmic Microwave Background (CMB) anisotropies due to Planck \cite{Ade:2015xua}, one can recognise that the conjecture is in obvious conflict with $\Lambda$CDM. In this note we assume the conjecture is correct and attempt a reconciliation with $\Lambda$CDM. To this end there are two possibilities: either a deviation from $\Lambda$CDM happens in the past at positive redshift $( z> 0 )$, so that it is observable and of interest to science, or it happens in the future $(z < 0)$ and is the subject of science fiction. In light of the recent tension in the Hubble constant $H_0$ between its local value $H_0 = 73.48 \pm 1.66$ km s$^{-1}$ Mpc$^{-1}$ \cite{rie18} and the Planck result  based on $\Lambda$CDM $H_0 = 66.93 \pm 0.62$ km s$^{-1}$ Mpc$^{-1}$ \cite{Aghanim:2016yuo}, it is natural to entertain the prospect that some departure happens at positive redshift. See \cite{Chen:2016uno,Cardona:2016ems,Kumar:2017dnp, Sola:2017jbl, DiValentino:2017zyq, DiValentino:2017iww, Sola:2017znb, Follin:2017ljs, Addison:2017fdm, Dhawan:2017ywl, DiValentino:2016hlg, DiValentino:2017rcr, Qing-Guo:2016ykt, Zhao:2017cud, Capozziello:2018jya, Barenboim:2017sjk, Mortsell:2018mfj, Feeney:2018mkj, Poulin:2018zxs, Bringmann:2018jpr, Kumar:2018yhh, Dumin:2018bjx} for attempts to explain the tension. 

The question then is how will this deviation from $\Lambda$CDM manifest itself? In this letter, we provide an alternative late-time cosmology. First, note that at low redshift, $z<2$, physics is dominated by only two parameters from the six of $\Lambda$CDM: the Hubble constant $H_0$ and matter density $\omega_{m}$ at redshift $z=0$. While $\Lambda$CDM leads to a Hubble parameter $H(z)$ satisfying $H^\prime(z)>0$ everywhere, $\Lambda$CDM in the Swampland suggests otherwise: an unstable de Sitter state giving rise to a turning point $H^\prime(z_*)=0$ at some low redshift $z_*$. In the most dramatic scenario, $H(z)$ diverges in the distant future, driven by $w<-1$ \cite{kam96}.  

Here we consider a model {with turning point at a critical redshift $z_*=z_*(\omega_m)$} that alleviates the $H_0$ tension \cite{vanPutten:2017bqf}. Governed by the exact same parameters $(H_0,\omega_m)$ as late-time $\Lambda$CDM, it allows a direct comparison with $\Lambda$CDM against current data. A key prediction of this model is the existence of a turning point in $H(z)$ at small {\em positive} $z_*$, here proposed as a new observable. As such, it is easily identified or falsified by upcoming experiments. 
Importantly, this potential outcome is consistent with current data, some of which already suggest $H(z)$ is constant at small $z$  \cite{Anderson:2013zyy, Cuesta:2015mqa} with a discernible preference for an equation of state $w < -1$ \cite{DiValentino:2016hlg, Qing-Guo:2016ykt, Zhao:2017cud, Capozziello:2018jya}. This observational trend is difficult to ignore. Somewhat understandably, since the Null Energy Condition (NEC) is violated \cite{Carroll:2003st, Cline:2003gs, Dubovsky:2005xd}, there are fewer theoretical models explaining such data, for example \cite{Csaki:2004ha,Csaki:2005vq,Sahlen:2005zw}, making our holographic model one of the few games in town. 

{\em Dark energy -}
The Swampland conjecture not only finds support in $H_0$ tension, but the status of one closely mirrors the other. On one hand, there is no proof that de Sitter vacua inhabit the Swampland, while on the other, the data supporting $H_0$ tension is inconclusive. Thus, on its own neither is satisfactory, but taken together, they  point to a future $(-1<z<0)$ distinct from the
de Sitter state predicted by $\Lambda$CDM. 

The simplest deviation from $\Lambda$CDM is to make dark energy dynamical. A common approach is to make the equation of state $w$ redshift dependent \cite{Qing-Guo:2016ykt, Zhao:2017cud}, but this introduces an extra parameter, thus making it easier to fit data. Instead, we begin with a physical argument based on holography and identify a minimal model with the same number of parameters as $\Lambda$CDM at late times. For simplicity we set Newton's constant and the velocity of light equal to 1, leaving $M_p^2=1/8\pi$. 

To get oriented, we recall that late-time cosmology of $\Lambda$CDM is governed by two constants: $H_0$ and $\omega_m$, the combined baryonic and cold dark matter (CDM) density today ($z=0$). Assuming pressure due to matter is zero, $p_m = 0$, one can solve the continuity equation for matter density $\rho_m$, before substituting back into the remaining Einstein equation to find an analytic solution for the normalised Hubble parameter $h(z) \equiv H(z)/H_0$, 
\begin{eqnarray}
h_{\textrm{$\Lambda$CDM}}(z) = \sqrt{1 - \omega_{m} + \omega_{m} (1+ z)^3}.
\label{EQN_LCDM}
\end{eqnarray}
Note, in deriving this expression, we have made use of the following relation between time and redshift, 
\be
\label{tz}
\frac{d}{d t}  = - (1+z) H \frac{d}{d z},  
\ee
thus allowing us to switch between time and the more natural parameter relevant for astronomy.  

{\em Turning point -}
While $\Lambda$CDM is one possibility for the normalised Hubble parameter, assuming analyticity at $z=0$, one can Taylor expand:
\begin{eqnarray}
h(z) = 1 + (1 + q_0) z + \frac{1}{2} (Q_0 + q_0 (1+q_0) )z^2 + b_3 z^3 \dots
\label{EQN_c} 
\end{eqnarray}
with coefficients expressed in terms of the values $(q_0, Q_0)$ at $z=0$ of the deceleration parameter $q(z)$ and its derivative $Q(z)$ 
(similar but not identical to the jerk parameter $j$)
\be
\label{qQ}
q(z) = -1 + (1+z) H^{-1} H'(z), \quad Q(z) = q'(z). 
\ee
This cubic polynomial provides a minimal model-independent setting to capture $(H_0,q_0,Q_0)$ of an analytic cosmological evolution. 

Simply differentiating (\ref{EQN_LCDM}) one can quickly confirm that $\Lambda$CDM only permits a constant value of $h(z)$ at asymptotic infinity, namely $z=-1$, so there is no turning point. Returning to our cubic polynomial, for small $z$ a turning point exists at 
\be
z_{*} \simeq - \frac{(1+q_0)}{Q_0 + q_0 (1+q_0)}. 
\ee
In contrast to $\Lambda$CDM, for models with a turning point at positive $z$, which are thus observable, assuming $Q_0 > 0$, we require $q_0 < -1$. We turn our attention to a model with just such a feature.  

{\em Holographic model -}
Our model is defined by a dynamical cosmological constant of the form \cite{van15}
\be
\label{DE}
\Lambda = (1 - q) H^2, 
\ee
where $q$ is the deceleration parameter (\ref{qQ}). The idea then is to simply replace $\Lambda$ with this expression in the Friedmann equations and resolve. Doing so, consistency of (\ref{DE}) with the Einstein equations demands that we introduce a negative ``dark pressure", which takes the form $p = H^2M_p^2 q$.  This may be attributed to the cosmological vacuum being dispersive: super-horizon scale fluctuations $(\omega < \omega_0$) acquire imaginary wave numbers, which gives rise to negative pressures $p$ - a hallmark property of dark energy. Our model has a number of similarities with Li's holographic model \cite{Li:2004rb}. Our independent analysis agrees that the relevant modification should be of order $H^2$ \cite{Wang:2016och} (based on arguments presented in \cite{Cohen:1998zx}), but there is no free constant parameter, so it is more constrained and therefore falsifiable. A further important difference is our apparent horizon is a past horizon, which makes the turning point observable. 

These comments aside, this leaves us with the task of solving the first Friedmann equation, which under the assumption that there is usual matter density, reduces to a single ODE: 
\be
\label{ODE} y'(z) = 3 (1+z)^2 \omega_m e^{-2 y} - \frac{1}{1+z}. 
\ee
where we have introduced $y(z) = \log h(z)$. The ODE permits an exact solution \cite{vanPutten:2017bqf}: 
\be 
h_{DDE}(z) = \frac{\sqrt{1+ (6/5) \omega_m [(1+z)^5-1]}}{1+z}.  
\label{EQN_h}
\ee
It is worth noting that there are still two constants, thereby facilitating direct comparison with $\Lambda$CDM. We stress that the model only depends on two parameters, so it should not be regarded as a replacement for $\Lambda$CDM, but simply a model offering contrasting behaviour at low redshift $z<2$. Despite not being a complete cosmology, it appears to perform well \cite{vanPutten:2017bqf} when confronted with cosmological measurements of $H(z)$ \cite{Farooq:2016zwm}.

Figure 1 illustrates this cosmological evolution alongside that of $\Lambda$CDM in the same two parameters.
It highlights a common epoch of matter dominated evolution ($q=1/2$, total pressure zero) yet distinct
behavior towards the present and future. For this model the associated dark energy equation of state between pressure $p_{\Lambda}$ and energy density $\rho_{\Lambda}$, $p_{\Lambda} = w \rho_{\Lambda}$, may be expressed as 
\be
w = \frac{2 q-1}{1-q}, 
\ee
which is in contrast to the $w = -1$ value in $\Lambda$CDM. 

\begin{figure}
\begin{center}
\includegraphics[scale=0.49]{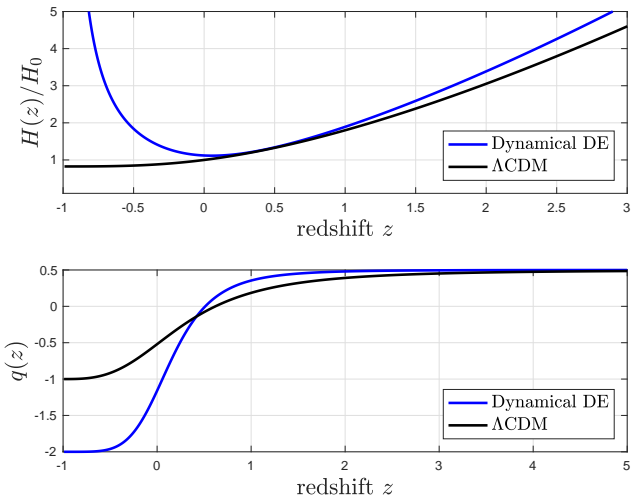}
\end{center}
\caption{Normalised cosmological evolution according to (\ref{ODE}) and $\Lambda$CDM (\ref{EQN_LCDM}). 
While evolution are relatively similar in the past ($z>0$), they are vastly distinct in the future ($z<0$).}
\label{fH}
\end{figure}

{\em Input from data -}
We now have an alternative cosmology based on a holographic argument, which provides a concrete example of a model with a turning point.  As stated, the current status of the data is inconclusive. That being said the tension between local measurements of $H_0$ \cite{rie18} and Planck analysis based on $\Lambda$CDM \cite{Aghanim:2016yuo} is difficult to ignore. Our goal here is simply to fix the two free parameters in the dynamical dark energy model using the existing tentative data. To that end, we import measurements of the Hubble parameter $H(z)$ \cite{Farooq:2016zwm}, which we restrict to $z < 2$, and use nonlinear regression from the MATLAB toolbox (fitnlm) to extract best-fit values for $H_0$ and $\omega_m$:
\be
H_0 = 74.9\pm 2.6 \,\mbox{km}\,\mbox{s}^{-1}\mbox{Mpc}^{-1}, \quad \omega_m = 0.2719\pm0.028. 
\label{EQN_om}
\ee
Note that the turning point provides a conceptually simple way to alleviate the tension in $H_0$ and favours the higher value $H_0=73.48 \pm 1.66\,$km\,s$^{-1}$Mpc$^{-1}$ from surveys of the local Universe \citep{rie18}. This is certainly an agreeable outcome. However, given an explicit model with an exact solution (\ref{EQN_h}), we can give a prediction for the turning point (Figure 2)
\begin{eqnarray}
z_* = \left( \frac { 5-6\omega_m } { 9 \omega_m } \right)^\frac{1}{5} -1 
\label{EQN_zs}
\end{eqnarray}
with the property that $z_* =0$ for $\omega_m=1/3$. Evaluated for (\ref{EQN_om}) gives
\be
z_* = {0.07 \pm 0.03}. 
\ee

\begin{figure}
\begin{center}
\includegraphics[scale=0.40]{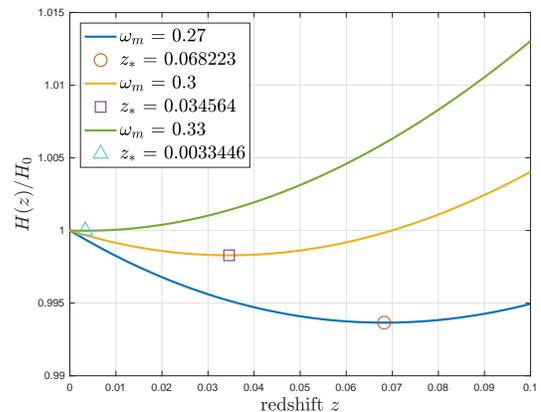}
\end{center}
\caption{Cosmological evolution outside the Swampland is expected to violate NEC ($w<-1)$, inevitably giving rise to a turn $z_*$ point in $H(z)$. Holography predicts $0<z_*<0.1$ as a new observable accessible by homogeneous supernova surveys covering extremely low redshifts of the Local Universe. Shown is the predicted location (\ref{EQN_zs}) for various canonical values of $\omega_m$, where $z_*$ for $\omega_m=0.27$ is illustrative for the anticipated location suggested by current cosmological data $H(z)$.}
\label{fH}
\end{figure}

The NEC $T_{ab} \nu^a\nu^b \ge 0$ for all null-vectors $\nu^b$ imposes
the condition $\rho + p \ge 0$, where $p$ is the dark pressure, {and hence} 
\citep{rub14} $H^\prime(z) \ge 0$. We see that monotonically increasing $H(z)$ is consistent with the NEC, but when it is decreasing across a turning point near $z=0$, 
the NEC is violated, in line with expectations. This corresponds to $w < -1$, which places it at odds with the bound $1+w \geq 0.15c^2$ \cite{Agrawal:2018own}. The fact that the NEC is violated tells us that we cannot expect to embed this model in a quintessence model. Indeed, attempts to do so lead to the scalar $\phi$ becoming complex before the turning point in the Hubble parameter is reached.

{\em Discussion -}
We have introduced a turning point in the Hubble parameter as a diagnostic of new physics at low positive redshift. This is a beguilingly simple resolution to the tension between Planck analysis of CMB data based on $\Lambda$CDM \cite{Ade:2015xua} and local measurements of the Hubble constant \cite{rie18}. Currently, cosmological measurements of the Hubble parameter are inconclusive, but the status of the data will improve rapidly in the coming years, thus making the presence of a turning point in $H(z)$ easily falsifiable. 

Assuming the tension persists, this also certainly rules out the $\Lambda$CDM cosmological model and along with it the de Sitter vacuum at future asymptotic infinity. In this sense, $H_0$ tension appears to be supportive of the Swampland conjecture that de Sitter vacua should be ruled out. That being said, persistent $H_0$ tension may present a double-edged sword and already a number of model-independent studies \cite{DiValentino:2016hlg, Qing-Guo:2016ykt, Zhao:2017cud, Capozziello:2018jya} favour a dark energy equation of state $w < -1$. This points to a violation of the NEC and a breakdown in EFT, with obvious implications for the Swampland program. 

Falling outside of the scope of good EFTs, it is understandable that dark energy models with $w < -1$ are poorly studied: theorists have a preference for EFT in spite of the Planck result, which points to a central value in this regime. Here, we presented a dynamical dark energy with a holographic component, providing a concrete example of a model with a turning point at redshift $z_* = 0.07 \pm 0.03$. The model is described by two free parameters, $H_0$ and $\omega_m$, thus allowing an ``apples with apples" comparison with $\Lambda$CDM at late times. 

This begs the question again is the tension real? As highlighted by the recent attention to the problem of stability of the de Sitter
 state of cosmology and arguments presented here, the importance of observations
 resolving the $H_0$ tension problem cannot be overstated in identifying the 
 physical nature of late-time cosmology.
 Our considerations favour the relatively high value of the Hubble parameter from 
 surveys of the Local Universe in significant departure from $\Lambda$CDM.
Independent confirmation may be established by observations of double neutron
star mergers \cite{gui17} from the current LIGO O3 run. 

\section*{Acknowledgements}
We thank S. Brahma, Md. W. Hossain, N. Kim, M. M. Sheikh-Jabbari and C. F. Uhlemann for discussion. We thank A. Shafieloo for constructive discussions on Taylor expansions. This research is supported in part by the National Research Foundation of Korea (No. 2015R1D1A1A01059793, 2016R1A5A1013277 and 2018044640) and in part by National Natural Science Foundation of China, Project 11675244. This research was also facilitated by the Ministry of Science, ICT \& Future Planning, Gyeongsangbuk-do and Pohang City.

\end{document}